\documentclass[aps,reprint,floatfix,twocolumn,superscriptaddress]{revtex4-1}

\usepackage{amsfonts,amssymb,amsmath,amsthm,graphicx}
\usepackage{url}
\usepackage{bbm}
\usepackage[usenames,dvipsnames]{xcolor}
\usepackage{natbib}
\usepackage{setspace}
\usepackage{subfigure}
\usepackage{comment}
\usepackage[normalem]{ulem}
\usepackage{todonotes}
\usepackage{siunitx}
\usepackage{xcolor}

\usepackage[colorlinks=true]{hyperref}
\hypersetup{linkcolor=blue,urlcolor=blue,citecolor=blue}

\date{\today}
\begin{document}

\title{Domain walls and Dzyaloshinskii-Moriya interaction in epitaxial Co/Ir(111) and Pt/Co/Ir(111)}

\author{Marco Perini}
\email{mperini@physnet.uni-hamburg.de}
\affiliation{Institute of Applied Physics, University of Hamburg, Jungiusstrasse 11, D-20355 
Hamburg, Germany}

\author{Sebastian Meyer}
\affiliation{Institute of Theoretical Physics and Astrophysics, Christian-Albrechts-Universit{\"a}t zu Kiel, Leibnizstrasse 15, D-24098 Kiel, Germany}

\author{Bertrand Dup\'{e}}
\altaffiliation{Present address: Institute of Physics, Johannes Gutenberg Universit\"{a}t Mainz, Staudingerweg 7, D-55128 Mainz, Germany}
\affiliation{Institute of Theoretical Physics and Astrophysics, Christian-Albrechts-Universit{\"a}t zu Kiel, Leibnizstrasse 15, D-24098 Kiel, Germany}


\author{Stephan von Malottki}
\affiliation{Institute of Theoretical Physics and Astrophysics, Christian-Albrechts-Universit{\"a}t zu Kiel, Leibnizstrasse 15, D-24098 Kiel, Germany}
 
\author{Andr\'e Kubetzka}
\affiliation{Institute of Applied Physics, University of Hamburg, Jungiusstrasse 11, D-20355 
Hamburg, Germany}
 
\author{Kirsten von Bergmann}
\affiliation{Institute of Applied Physics, University of Hamburg, Jungiusstrasse 11, D-20355 
Hamburg, Germany}

\author{Roland Wiesendanger}
\affiliation{Institute of Applied Physics, University of Hamburg, Jungiusstrasse 11, D-20355 
Hamburg, Germany}

\author{Stefan Heinze}
\affiliation{Institute of Theoretical Physics and Astrophysics, Christian-Albrechts-Universit{\"a}t zu Kiel, Leibnizstrasse 15, D-24098 Kiel, Germany}

 \begin{abstract}
We use spin-polarized scanning tunneling microscopy and density functional theory (DFT) to study 
domain walls (DWs) and the Dzyaloshinskii-Moriya interaction (DMI) in epitaxial films of Co/Ir(111) and Pt/Co/Ir(111).
Our measurements reveal DWs with fixed rotational sense on one monolayer of Co on Ir, with a wall width around 2.7~nm. 
With Pt islands on top, we observe that the DWs occur mostly in the uncovered Co/Ir areas, 
suggesting that the wall energy density is higher in the Pt/Co/Ir(111).
From DFT we find an interfacial DMI that stabilizes N\'{e}el-type DWs with clockwise rotational sense. 
The calculated DW widths are in good agreement with the experimental observations. 
The total DMI nearly doubles from Co/Ir(111) to Pt/Co/Ir(111), however, 
in the latter case the DMI is almost entirely due to the Pt with only a minor Ir contribution. 
Therefore a simple additive effect, where both interfaces contribute significantly to the total DMI, is not observed for one atomic Co layer.

 \end{abstract}
\maketitle

In the field of spintronics, 
localized non-collinear magnetic structures such as domain walls (DWs) and skyrmions are
promising candidates for innovative technological applications \cite{Parkin190, Fert2013}.
A key aspect in this field of research is the right
choice of materials which can form such magnetic structures. 
Recently, cobalt-based multilayer systems
have gained significant attention 
due to the observation of magnetic skyrmions at room temperature
\cite{Moreau-Luchaire2016, Boulle2016, Pulecio2016, Soumyanarayanan2017}.
The stabilization of these non-collinear magnetic states 
requires the competition of different interactions 
\cite{Moreau-Luchaire2016, Boulle2016, Pulecio2016, Soumyanarayanan2017, Heinze2011, Bogdanov1989, Bogdanov:2001aa}.
In particular, 
the interfacial Dzyaloshinskii-Moriya interaction (DMI)
\cite{Dzyaloshinskii1957,Moriya1960,Crepieux1998,Bode2007} 
favors magnetic structures with fixed rotational sense,
such as N\'{e}el-type DWs \cite{Heide2008, Thiaville2012} and 
skyrmions \cite{Bogdanov1989, Bogdanov:2001aa,Heinze2011,Romming2013}.
It occurs in systems with large spin-orbit coupling (SOC) and broken inversion symmetry,
e.g. at the interfaces of cobalt with heavy materials possessing large SOC,
such as iridium or platinum \cite{Fert1990}.
The DMI is extremely sensitive to the interface quality \cite{Wells2017}.
Sputtered films on amorphous or polycrystalline substrates, 
as those studied in Refs.~\onlinecite{Moreau-Luchaire2016, Boulle2016, Pulecio2016}, 
are not well suited for a microscopic understanding of the details of the DMI
since the interface roughness of such films is difficult to characterize experimentally 
and hard to include in first-principles calculations. 
Ab-initio methods are also crucial to investigate the behavior of the DMI
in a multilayered system \cite{Dupe2016},
where multiple interfaces contribute to the total DMI
and the effects of each of them are hard to disentangle experimentally.

Epitaxially grown ultrathin films on single crystal substrates are best suited for a detailed study of interface DMI. 
These model-type systems can also be compared to first-principles calculations, 
enabling a disentanglement of the relevant interactions involved in the
formation of magnetic states.
Previous works have reported
opposite signs of DMI at the Co/Ir and Co/Pt interfaces \cite{Dupe2014, Yang2015, Vida2016},
and predicted an enhancement of the total DMI 
when the Co is sandwiched between the two heavy metals in
a multilayer configuration, an effect often referred to as additive DMI \cite{Moreau-Luchaire2016,Yang2016}.
This effect is not always observed in experiments with epitaxial films,
as shown in Ref. \onlinecite{Chen2015} for Co/Ir/Ni multilayers.
This suggests that the mechanism of additive DMI still requires a deeper understanding.
Magnetic DWs have been used in different materials to determine the sign of the 
interfacial DMI \cite{Chen2015, Corredor2017, Meckler2009, Hrabec2014}.
For DWs with a width in the single digit nanometer range, spin-polarized 
scanning tunneling microscopy (SP-STM) has the necessary spatial resolution to image them.
A previous SP-STM study on Co/Ir(111) showed that
isolated Co monolayer (ML) islands on Ir are ferromagnetic in the out-of-plane direction,
with large coercive fields $H_c$ \cite{Bickel2011}. 
Such islands grow pseudomorphically on the Ir substrate, 
making Co/Ir(111) an optimal model system to 
study the properties of the interfacial DMI both experimentally and theoretically.

Here, we study pseudomorphic films of a Co ML and
an atomic bilayer of Pt/Co on an Ir(111) surface. 
Using SP-STM we observe that the films exhibit out-of-plane magnetic domains,
separated by nanometer-wide DWs. 
The DW widths are in very good agreement with those calculated from DFT parameters. 
We demonstrate experimentally that the DWs have a unique rotational sense. 
DFT calculations reveal that the DMI is almost twice as large
in Pt/Co/Ir(111) as in Co/Ir(111).
Surprisingly, the DMI in Pt/Co/Ir(111) is dominated by Pt,
while the Ir contribution is nearly quenched compared to the bare Co/Ir(111) system. 
For a sandwich structure with a Co double layer,
on the other hand, the DMI further increases and 
both Pt and Ir contribute significantly, 
leading to an additive DMI effect as suggested in Ref.~\onlinecite{Moreau-Luchaire2016}.

A typical sample of Co grown on the Ir(111) substrate is shown in
Fig.~\ref{fig: fig1}(a) (see Supplemental Material for experimental details \cite{Supplement}), in which the topography is
colorized with the measured spin-polarized differential tunneling 
conductance ($\mathrm{d}I/\mathrm{d}U$), 
obtained using a magnetic tip sensitive to the out-of-plane component of
the sample magnetization. 
Despite the 8\% lattice mismatch,
the Co ML mainly grows pseudomorphically with occasional dislocation lines to
reduce the strain. Due to the smooth step flow growth we assume that the Co is fcc-stacked. 
The two levels of contrast on the Co indicate areas with a
magnetization pointing in opposite out-of-plane directions \cite{Bickel2011},
see sketched arrows in Fig.~\ref{fig: fig1}(a).

\begin{figure}
\centering
\includegraphics[scale=1.08]{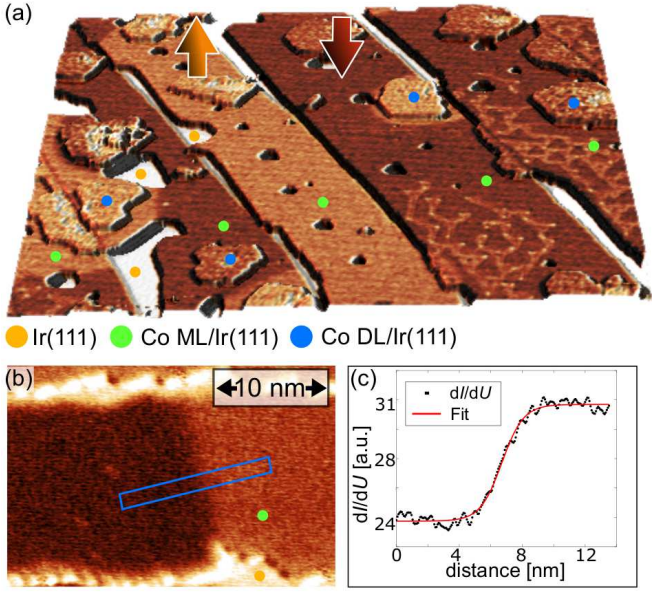}
\caption{(a) Perspective view of the topography of about 1.2 atomic layers of Co on Ir(111)
(200 nm $\times$ 180 nm), 
colorized with the simultaneously measured spin-resolved $\mathrm{d}I/\mathrm{d}U$ signal. 
The arrows indicate 2 oppositely magnetized domains of the out-of-plane ferromagnetic Co stripes.
The colored dots identify the different areas of Ir, Co monolayer (ML) and double layer (DL).
(b) Closer view of a single domain wall.
The signal inside the blue box is averaged in the short direction and
then plotted against the long direction in (c). 
The red line is a fit of the data points to Eq. \eqref{eq:Tanh}.
(a): $U = -580 \,\text{mV}$, $I = 500\,\text{pA}$, $T = 8 \,\text{K}$; 
(b): $U = -450\,\text{mV}$, $I = 500 \,\text{pA}$, $T = 8 \,\text{K}$.
}
\label{fig: fig1}
\end{figure}

Fig.~\ref{fig: fig1}(b) shows a d$I$/d$U$~map of a DW on a ML Co stripe, which separates oppositely magnetized out-of-plane domains, 
and the $\mathrm{d}I/\mathrm{d}U$ signal is plotted in 
Fig.~\ref{fig: fig1}(c) as a function of the position across the wall.
The DW profile is fitted according to 

\begin{equation}
y = y_0 + y_{sp}  \text{tanh}\left(\frac{x-x_0}{w/2}\right) \label{eq:Tanh}
\end{equation}

\noindent where $y_0$ and $y_{sp}$ are the spin averaged and
spin-polarized contributions to the $\mathrm{d}I/\mathrm{d}U$ signals, respectively,
$x_0$ is the center of the DW, and $w$ is the wall width.
We fitted 8 walls from different samples, 
obtaining an average of $w = 2.7 \pm 0.3 \,\text{nm}$, 
where the uncertainty represents the standard deviation of the values.

\begin{figure}
\centering
\includegraphics[scale=1.05]{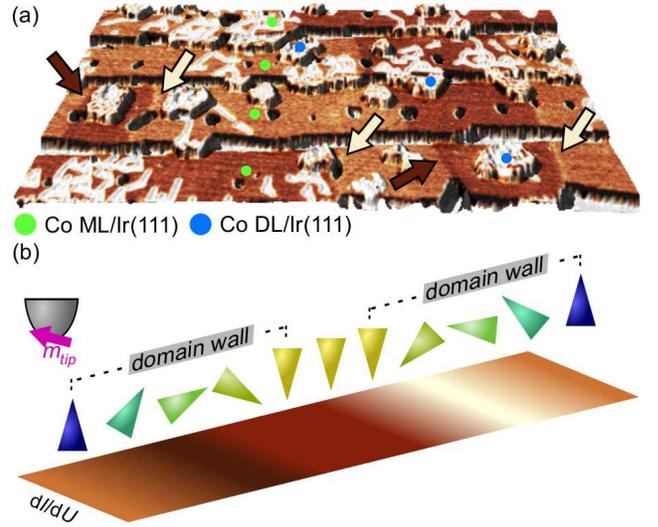}
\caption{(a) 3D view of Co ML terraces (280 nm $\times$ 210 nm) colorized with 
the simultaneously measured spin-resolved $\mathrm{d}I/\mathrm{d}U$ signal. The tip has a canted magnetization, i.e.\ it is
sensitive to both in-plane and out-of-plane magnetization components. 
The arrows indicate the position of the DWs and are colorized according to the $\mathrm{d}I/\mathrm{d}U$ contrast of the walls.
Neighboring walls in a stripe always show an alternating contrast.
The colored dots identify the different areas of Co ML and DL.
(b) Schematics of spin-resolved $\mathrm{d}I/\mathrm{d}U$ contrast 
across two consecutive DWs with a tip that has a canted magnetization. 
(a): $U = -450\,\text{mV}$, $I = 500\,\text{pA}$, $T = 8 \,\text{K}$.
}
\label{fig: fig2}
\end{figure}

While Fig.~\ref{fig: fig1}(b) shows a DW imaged using a fully out-of-plane magnetic tip, 
Fig.~\ref{fig: fig2}(a) shows several DWs observed with a canted magnetic tip, 
which also has a significant in-plane component of the magnetization.  
With such a tip the DWs appear as bright or 
dark stripes depending on the magnetization direction within the walls. We find a correlation of the magnetization within a wall with the order of the out-of-plane orientation of the domains they separate,
as indicated by the differently colorized arrows in Fig.~\ref{fig: fig2}(a)).
This strict sequence of spin-resolved d$I$/d$U$~contrasts demonstrates that the DWs in the Co ML on Ir(111) have a unique rotational sense, as sketched in Fig.~\ref{fig: fig2}(b). 
This is confirmed by investigating 13 independent 
DWs with the same magnetic tip (not shown). 
We conclude that the interfacial DMI is large enough to select a unique rotational sense and 
propose that it also induces N\'{e}el-type DWs according to the symmetry selection rules~\cite{Heide2008, Thiaville2012, Bergmann2014}.
An attempt to measure the absolute rotational sense of the DWs 
is presented in the Supplemental Material~\cite{Supplement}.

\begin{figure}
\centering
\includegraphics[scale=1.1]{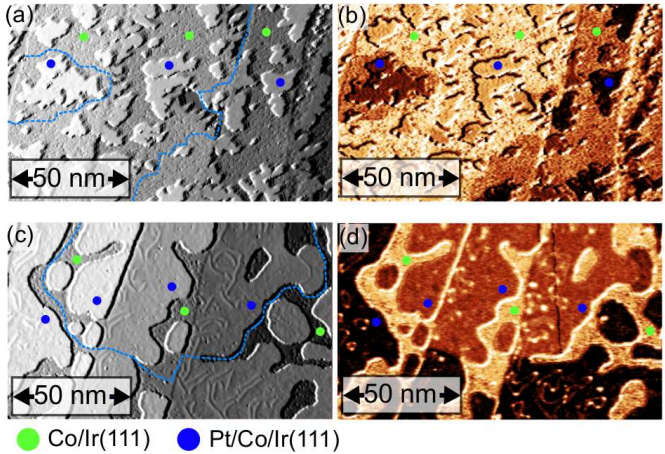}
\caption{(a),(c) Superposition of constant-current topography and 
current map of Pt/Co/Ir(111) in the as-grown case and after post-annealing, respectively. 
(b),(d) Corresponding $\mathrm{d}I/\mathrm{d}U$ maps with an out-of-plane magnetic tip.
The dashed blue line traces the position of the domain walls, 
which avoid Pt/Co when possible.
(a), (b): $U = -450 \,\text{mV}$, $I = 1 \,\text{nA}$, $T = 8 \,\text{K}$; 
(c), (d): $U = -550 \,\text{mV}$, $I = 1 \,\text{nA}$, $T = 8 \,\text{K}$.
}
\label{fig: fig3}
\end{figure}

To investigate the effect on the DWs when the Co is sandwiched between Pt and Ir,
we performed additional measurements on ML Pt islands on top of an almost
complete ML of Co on Ir(111). 
Fig.~\ref{fig: fig3}(a) shows pseudomorphic Pt islands 
with irregular shapes grown with the Co/Ir(111) held at room temperature. 
The spin-resolved d$I$/d$U$~contrast of the Pt islands in Fig. \ref{fig: fig3}(b) 
is strongly correlated with that of the surrounding Co ML, and 
we conclude that the Co induces a magnetic polarization in the Pt and
that the Pt/Co atomic bilayer is also out-of-plane ferromagnetic. 
A closer look to the data in Fig.~\ref{fig: fig3}(b) reveals two slightly different d$I$/d$U$~contrast levels for each magnetization direction,
originating from islands grown in the two different stackings.
In order to obtain more extended films of Pt, 
we post-annealed the samples at moderate temperature ($T \simeq $ \ang{500}C) after the Pt deposition.
This results in a mostly fcc-stacked Pt layer attached to step edges
as shown in Fig.~\ref{fig: fig3}(c). 
The magnetic domain structure of this sample is evident from the simultaneously aquired d$I$/d$U$~map in Fig. \ref{fig: fig3}(d).
The Pt layer still presents out-of-plane magnetization, 
induced by the Co ML (see Fig. \ref{fig: fig3}(d)).
A close analysis of the position of the DWs for these two differently prepared samples,
see blue lines in Figs.~\ref{fig: fig3}(b) and \ref{fig: fig3}(d), demonstrates
that the DWs tend to avoid the Pt/Co layer if possible,
increasing their length in order to remain on the bare Co. 
This suggests a higher domain wall energy in the Pt/Co bilayer than in the Co ML on Ir(111). 
We were able to extract the width $w$ of only 2 DWs in the Pt/Co layer, 
due to their tendency to sit in very small constrictions,
obtaining the values of 2.7 and 3.6 nm, similar to the ones in Co/Ir(111). 
To minimize the effect of the constriction size on the DW width \cite{Bruno1999}, 
we have selected walls occurring in areas at least twice as large as $w$.

\begin{figure*}
\includegraphics[scale=0.75]{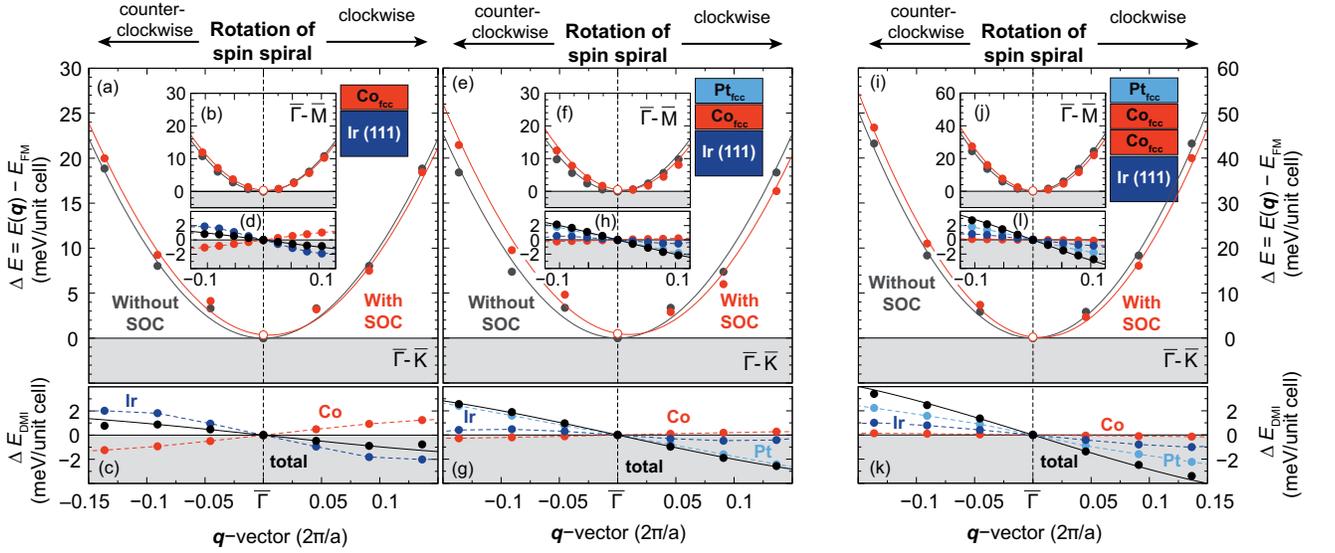}
\caption{(a), (b) Calculated energy dispersion \(E(\mathbf{q})\) of flat, cycloidal
spin spirals for Co/Ir(111) ((e), (f) for
Pt/Co/Ir(111), (i), (j) for Pt/Co/Co/Ir(111)) without (grey dots) and with (red dots)
spin orbit interaction for left and right rotating spin spiral states in
\(\overline{\Gamma}\)-\(\overline{\mathrm{K}}\) and
\(\overline{\Gamma}\)-\(\overline{\mathrm{M}}\) direction, respectively.
The grey line is a fit of the dispersion to the Heisenberg model, 
the red line is a fit including the DMI and the magnetocrystalline anisotropy. 
(c), (d), (g), (h), (k), (l) Element-resolved energy contribution for Co/Ir(111),
Pt/Co/Ir(111) and Pt/Co/Co/Ir(111)
to the DMI \(\Delta E_{\text{DMI}} (\mathbf{q})\).
Note that the total DMI is the sum of all contributions
and that all energies are given per unit cell (i.e. two magnetic atoms for Pt/Co/Co/Ir(111)).
The black curve is the fit of the DMI.
All fits are made within the nearest-neighbor approximation.}
\label{Fig: energy dispersion}
\end{figure*}

\begin{table*}
\centering
\caption {Calculated values for the nearest neighbor exchange interaction \(J\) (meV),
Dzyaloshinskii-Moriya interaction \(D\) (meV),
magnetocrystalline anisotropy energy \(K\) (meV), 
calculated energy difference \cite{Heide2008} 
$\Delta E_{\text{DW}-\text{FM}} = \frac{4}{a} \sqrt{2 \vert{JK}\vert} - \frac{2}{a} \pi \sqrt{3} \vert{D\vert}$
between a DW and 
the FM state (meV/nm), and calculated DW width
\(w = 2a\sqrt{{3 J} \over { 2 \vert K \vert}}\), 
with \(a\), the lattice constant of the (111) plane of the substrate, compared to the experimental DW width $w_\text{exp}$ (nm). 
$D < 0$ represents clockwise rotation and
\(K < 0\) represents an out-of-plane easy magnetization axis.
All magnetic interactions are given per Co atom.}
\begin{ruledtabular}
\begin{tabular}{lcccccc} 
System & \(J\) &  \(D\) & \(K\) & \(\Delta E_{\text{DW}-\text{FM}}\) & \(w\) & $w_\text{exp}$\\ \hline 
Co/Ir(111) & \(+17.6\) & \(-0.54\) & \( -0.73 \) & \(+53.3\) & \(3.2\) & $2.7 \pm 0.3 $\\
Pt/Co/Ir(111) & \(+18.0\) & \(-1.12\) & \( -1.00 \) & \(+43.8\) & \(2.8\) & $2.7, 3.6$ \\
Pt/Co/Co/Ir(111) & $ +20.0 $ & $-0.80 $ & $-0.09 $ & $ -4.1$ & $ 9.9 $ & $ - $
\end{tabular}
\end{ruledtabular}
\label{Tab: Values}
\end{table*}

To investigate the effects of the Pt/Co and Co/Ir interfaces on the total DMI, 
we have performed density functional theory (DFT) calculations.
We apply the full-potential linearized augmented plane wave method (FLAPW)
\cite{Wimmer1981,Jansen1984} as implemented in the FLEUR code~\cite{FLEUR}. 
We have performed structural relaxations for films of Co/Ir(111) and Pt/Co/Ir(111), 
where both Co and Pt are single atomic layers,
and have considered fcc stackings of Co and 
Pt based on the experimental observations.
For both systems we calculated the energy dispersion \(E(\mathbf{q})\) of homogeneous,
flat spin spirals \cite{Kurz2004, Zimmermann2014} along 
the high symmetry directions \(\overline{\Gamma}\)--\(\overline{\mathrm{M}}\) and
\(\overline{\Gamma}\)--\(\overline{\mathrm{K}}\) of the two dimensional Brillouin zone.
To obtain the DMI,
we calculated the energy contribution due to DMI \(\Delta E_{\text{DMI}} (\mathbf{q})\)
in first order perturbation for every value of \(\mathbf{q}\) of the energy
dispersion \cite{Zimmermann2014, Heide2009, Meyer2017}
(see Supplemental Material for computational details \cite{Supplement}). 

We restrict ourselves to the region around the \(\overline{\Gamma}\)-point and
use the nearest-neighbor approximation for exchange and DMI, 
which are expressed by the parameters \(J\) and \(D\), respectively.
The magnetocrystalline anisotropy energy (MAE) is obtained by
constraining the spin quantization axis in out-of-plane (\(\perp\)) and 
in-plane (\(||\)) directions with respect to the film, 
determining the uniaxial anisotropy via \(K = E_{\perp} - E_{||} \).

Figs.~\ref{Fig: energy dispersion}(a) and \ref{Fig: energy dispersion}(b) show the calculated energy dispersion
for Co/Ir(111) in \(\overline{\Gamma}\)-\(\overline{\mathrm{K}}\) and
\(\overline{\Gamma}\)-\(\overline{\mathrm{M}}\) directions, respectively. 
The ferromagnetic (FM) state is the lowest in energy and the nearest-neighbor
exchange interaction \(J\) (Tab.~\ref{Tab: Values}) 
describes very well the energy dispersion.
\(\Delta E_{\text{SOC}} (\mathbf{q})\) is small compared to the variation of the exchange \(E(\mathbf{q})\)
(cf.~\(D\) in Tab.~\ref{Tab: Values}).
The interfacial DMI results as the sum of the individual contributions from the 
different layers
(see Fig. \ref{Fig: energy dispersion}(c) and \ref{Fig: energy dispersion}(d)), and
it prefers clockwise rotating spin spirals ($D < 0$).
The Ir surface dominates the total DMI, 
however Co has a significant contribution of opposite sign,
which we attribute to its small coordination number \cite{Meyer2017}.
The MAE prefers an out-of-plane magnetization ($K_\text{eff}$ = $-0.7\,\text{meV}$,
cf.~Tab.~\ref{Tab: Values}), that favors collinear states over spin spirals.

An atomic Pt overlayer does not modify the exchange interaction significantly with respect to Co/Ir(111)
(see Figs.~\ref{Fig: energy dispersion}(e), \ref{Fig: energy dispersion}(f) and Tab.~\ref{Tab: Values}). 
In contrast, the MAE increases by about 50\% and 
the total DMI is more than twice larger,
as expected from an additive DMI effect due to the two interfaces with Pt and Ir \cite{Moreau-Luchaire2016}.
However, Figs.~\ref{Fig: energy dispersion}(g) and \ref{Fig: energy dispersion}(h) show that the DMI
is dominated by the contribution from the Pt overlayer. 
The Ir substrate has a minor contribution which is nearly 
quenched compared to its value in Co/Ir(111) (cf.~Fig.~\ref{Fig: energy dispersion}(c)).

Adding another Pt adlayer does not change significantly the total DMI nor the 
contributions from the two interfaces (see Supplemental Material \cite{Supplement}).
We conclude that the hybridization of the Co layer with both the Ir and
the Pt has a decisive impact on the resulting DMI.
A simple additive effect as suggested in Ref.~\onlinecite{Moreau-Luchaire2016},
where the Pt/Co and Co/Ir interfaces contribute with a similar DMI
strength in a Pt/Co/Ir multilayer,
does not apply for one atomic Co layer. 

In order to test our interpretation, 
we have performed calculations with an additional Co layer,
i.e.~Pt/Co/Co/Ir(111) (cf.~Figs.~\ref{Fig: energy dispersion}(i - l)).
While the exchange rises by about 10\%, 
the MAE is reduced by a factor of 10 compared to Pt/Co/Ir(111) (see Tab.~\ref{Tab: Values}).
The total DMI increases compared to the Co ML system, 
because of an enhanced Ir contribution,
whereas the Pt contribution is similar to the one in Pt/Co/Ir(111).
Thus for a Co double layer both the Pt/Co and Co/Ir interfaces, 
which are farther apart from each other, 
contribute significanlty to the total DMI, supporting the interpretation of 
an additive DMI effect of Ref.~\onlinecite{Moreau-Luchaire2016}.
Note that the value of $D$ given in Tab.~\ref{Tab: Values} for Pt/Co/Co/Ir(111) 
is still smaller than for Pt/Co/Ir(111) since it is given per Co atom.

The DW width and energy obtained with the DFT values of exchange, DMI, and MAE 
are given in Tab.~\ref{Tab: Values}. 
The DW widths obtained for Co/Ir(111) and Pt/Co/Ir(111) are 
in good agreement with the experimental values. 
Consistent with the spin spiral energy dispersions and
the experimental observations, 
DWs are energetically unfavorable ($\Delta E_{\text{DW}-\text{FM}} > 0$) 
for Co/Ir(111) and Pt/Co/Ir(111). 
The smaller DW energy in Pt/Co/Ir(111) seems to disagree with 
the experimental observation of DWs being more likely to occur on the Co ML. 
However, the difference between the 
two DW energies is small with respect to the accuracy of the parameters
(see Supplemental Material for details \cite{Supplement}). 

In conclusion, 
we have combined SP-STM measurements and DFT calculations to investigate
DWs and interfacial DMI on pseudomorphic
Co/Ir(111) and Pt/Co/Ir(111) ultrathin films.
We observe DMI-stabilized N\'{e}el walls with a fixed rotational sense,
separating out-of-plane magnetic domains. 
Our calculations show a clockwise sense of magnetization rotation across the walls.
The DMI increases by more than a factor of two in the sandwich structure. 
However, this increase is dominated by Pt with a negligible Ir contribution.
For a Co double layer between Pt and Ir both interfaces contribute significantly to the total DMI.
This highlights the importance of interlayer hybridization and
Co film thickness in determining the total value of the DMI in magnetic multilayers.

\begin{acknowledgments}
We gratefully acknowledge computing time at the supercomputer 
of the North-German Supercomputing Alliance (HLRN).
This project has received funding from the European Unions
Horizon 2020 research and innovation programme under grant agreement
No 665095 (FET-Open project MAGicSky), and from the DFG via SFB668-A8.
\end{acknowledgments}

\bibliography{literature}

\end{document}